\documentclass[%
 reprint,
 superscriptaddress,
 amsmath,amssymb,
 aps,
 prl,
]{revtex4-1}

\usepackage{graphicx}% Include figure files
\usepackage{dcolumn}% Align table columns on decimal point
\usepackage{bm}% bold math
\usepackage{hyperref}% add hypertext capabilities

\usepackage[bold]{hhtensor} % \matr, \vec

\usepackage{color}
\usepackage{suffix}
\usepackage{float}

\usepackage{amsmath}

\definecolor{colorstraight}{rgb}{0.25,0.5,0.75}
\definecolor{colortss}{rgb}{0.75,0.25,0.25}
\definecolor{colorus}{rgb}{0.75,0.75,0.25}
\definecolor{colorrtime}{rgb}{0.75,0.25,0.75}
\definecolor{colorre}{rgb}{0.25,0.25,0.25}
\definecolor{colorsym}{rgb}{0.25,0.5,0.75}
\definecolor{colorasym}{rgb}{0.25,0.75,0.5}
\definecolor{colorpassive}{rgb}{0.25,0.25,0.25}
\definecolor{colorflowhigh}{rgb}{0.25,0.5,0.75}
\definecolor{colorflowlow}{rgb}{0.15,0.4,0.65}

\newcommand{\NameStraight}{bottom-heavy}
\WithSuffix\newcommand\NameStraight*{\mathrm{b}\mbox{-}\mathrm{h}}
\newcommand{\NameTss}{surfer}
\WithSuffix\newcommand\NameTss*{\mathrm{surf}}

\newcommand{\norm}[1]{\left\lvert #1 \right\rvert}
\newcommand{\abs}[1]{\left\lvert #1 \right\rvert}

\newcommand{\ParticlePosition}{\vec{X}}
\newcommand{\FlowVelocity}{\vec{u}}
\newcommand{\FlowVorticity}{\vec{\omega}}

\newcommand{\Gradients}{\vec{\nabla} \FlowVelocity}
\WithSuffix\newcommand\Gradients*{\left( \vec{\nabla} \FlowVelocity \right)}

\newcommand{\Direction}{\hat{\vec{z}}}
\newcommand{\SwimmingVelocity}{V_{\text{swim}}}
\newcommand{\SwimmingDirection}{\hat{\vec{p}}}
\WithSuffix\newcommand\SwimmingDirection*{\vec{p}}
\newcommand{\SwimmingDirectionOpt}{\hat{\vec{p}}_{\NameTss*}}
\WithSuffix\newcommand\SwimmingDirectionOpt*{\vec{p}_{\NameTss*}}

\newcommand{\TimeHorizon}{\tau}
\newcommand{\TimeHorizonOpt}{\tau^*}
\newcommand{\Performance}{V_{\mathrm{eff}}}
\WithSuffix\newcommand\Performance*{V_{\mathrm{eff}}}

\newcommand{\Kolmogorov}{\eta}
\newcommand{\KolmogorovT}{\tau_{\Kolmogorov}}
\newcommand{\KolmogorovU}{u_{\Kolmogorov}}

\newcommand{\ReorientationTime}{\tau_{\mathrm{align}}}
\newcommand{\CorrelationTime}{\TimeHorizon_\mathrm{corr}}

\newcommand{\ControlDirection}{\hat{\vec{n}}}
\WithSuffix\newcommand\ControlDirection*{\vec{n}}
\newcommand{\ControlDirectionOpt}{\hat{\vec{n}}_{\NameTss*}}
\WithSuffix\newcommand\ControlDirectionOpt*{\vec{n}_{\NameTss*}}

\begin{document}

\preprint{APS/123-QED}

\title{Surfing on turbulence: a strategy for planktonic navigation}

\author{Rémi Monthiller}
\author{Aurore Loisy}
\affiliation{
    Aix Marseille Univ, CNRS, Centrale Marseille, IRPHE, Marseille, France
}
\author{Mimi A. R.  Koehl}
\affiliation{
    Department of Integrative Biology, University of California, Berkeley, CA 94720-3140, USA
}
\author{Benjamin Favier}
\author{Christophe Eloy}
\affiliation{
    Aix Marseille Univ, CNRS, Centrale Marseille, IRPHE, Marseille, France
}

\begin{abstract}
    In marine plankton, many swimming species can perceive their environment with flow sensors. 
    Can they use this flow information to travel faster in turbulence? To address this question, we consider plankters swimming at constant speed, whose goal is to move upwards. 
    We propose a robust analytical behavior that allows plankters to choose a swimming direction according to the local flow gradients. 
    We show numerically that such plankters can ``surf'' on turbulence and reach net vertical speeds up to twice their swimming speed. 
    This new physics-based model suggests that planktonic organisms can exploit turbulence features for navigation.
\end{abstract}

\maketitle

Plankton are small organisms drifting in oceans. 
While they are carried by the ambient turbulent flow, many can swim and are equipped with hair-like mechanosensory organelles used to sense flows relative to their bodies, i.e. velocity gradients \cite{lenz1997zooplankton, wheeler2019not, kiorboe1999hydrodynamic, yen1992mechanoreception}. 
Besides, many can sense gravity or light \cite{richards1996diel, adams1999phototaxis, long2004navigational}, both indicating which direction is up. 
Here we focus on a planktonic navigation problem in turbulence: can motile planktonic organisms use local hydrodynamic signals to travel faster than their swimming speed along the vertical direction?

Vertical migration is an important task for many types of plankton. 
For instance, copepods are abundant millimetric crustaceans that move upwards to food-rich surface waters at night and downwards away from visual predators during the day \cite{richards1996diel, ringelberg2009diel}. 
Various planktonic larvae migrate up or down into currents at particular depths that transport the larvae horizontally \cite{mcedward2020ecology, welch2001flood, kingsford2002sensory}. 
Some larvae, when ready to settle, sink or swim downwards in response to chemical cues \cite{hadfield2004rapid} or mechanical stimuli due to turbulence \cite{fuchs2004sinking, fuchs2015hydrodynamic, fuchs2018waves, wheeler2015isolating}.

The navigation task faced by plankters has two features: 
(1) plankters only sense local flow information; and
(2) plankters only sense velocity gradients, not flow velocities. 
This makes planktonic navigation different from the Zermelo's navigation problem \cite{zermelo1931navigationsproblem} (where agents sense the full velocity field) and the bird soaring problem \cite{Reddy2016a,Reddy2018} (where birds sense the vertical flow velocity).

Problems of planktonic navigation have been recently approached using reinforcement learning \cite{Colabrese2017,Gustavsson2017,Qiu2021aarxiv,Qiu2021barxiv,Alageshan2020,Gunnarson2021arxiv}.
These studies showed that strategies based on local gradients can be learned in simple flows. 
Training a microswimmer in unsteady 3D turbulence remains however challenging \cite{Alageshan2020,Qiu2021barxiv}.
Besides, the strategies learned are not necessarily optimal nor easily interpretable.

Different models of zooplankton in turbulence have explored the consequences of various behaviors \cite{michalec2020efficient, visser2009swimming}.
For example, models of slowly-swimming planktonic larvae of different shapes in turbulent flow \cite{koehl2007individual, koehl2015swimming, pujara2018rotations} 
or in shear \cite{grunbaum2003form, grunbaum2003form} have shown how steady swimming or sinking, or behavioral responses to chemical or hydrodynamic cues can affect where they are transported by the ambient flow. 
A model of copepods finding patches of prey in turbulence included sensory cues, but not transport by ambient water motion \cite{woodson2007prevalence}.  
These data-based models are however purely empirical.

\begin{figure}[b]
    \centering
    \includegraphics{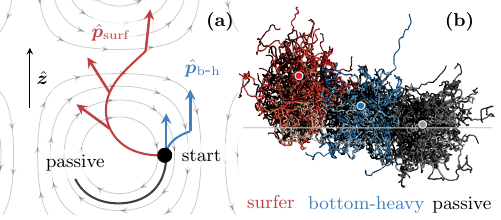}
    \caption{
        Plankters can exploit velocity gradients to ``surf'' on the flow: (a) 2D Taylor-Green vortex flow; (b) 3D turbulent flow. We compare the trajectories of surfers (red) to those of bottom-heavy swimmers (blue), which always swim upwards. We also show trajectories of passive particles (black). In (a), arrows show the swimming direction. In (b), the gray line shows the depth of the initial positions and circles show the average final vertical position for the same turbulent flow.
    }
    \label{fig:look}
\end{figure}

In this Letter we propose an approach based on physical principles. 
We model the navigation problem of going upwards and we derive an approximate solution, within well-defined hypotheses, where the response (preferred swimming direction) is an analytic function of the environmental signal (local velocity gradient).
This behavior can be interpreted as ``surfing'' on the flow (Fig. \ref{fig:look}):
to  exploit upward fluid motions, the plankter chooses a swimming direction by assuming that the flow is locally steady and linear.

We consider a plankter in homogeneous, isotropic turbulence. 
Its task is to go as fast as possible in a target direction, which is chosen to be $\Direction$, the vertical, without loss of generality.
We model the plankter as an active particle with position $\ParticlePosition (t)$, swimming in direction $\SwimmingDirection(t)$ at constant swimming speed $\SwimmingVelocity$ in a 3D flow velocity field $\FlowVelocity (\vec{x}, t)$. % of vorticity $\FlowVorticity (\vec{x}, t) = \vec{\nabla} \times \FlowVelocity$.
The plankter is assumed to be inertialess, neutrally buoyant, and small compared to the Kolmogorov scale $\eta$ (the scale of the smallest turbulent flow features \cite{frisch1995turbulence}). 
It actively controls its orientation by choosing a  direction $\ControlDirection$. 
We start by assuming that the swimming direction $\SwimmingDirection$ is always aligned with this chosen direction $\ControlDirection$. This corresponds to the limit of instantaneous reorientation (the effect of a finite reorientation time will be addressed below).
Under these assumptions, the equations of motion are
\begin{subequations}\label{eq:motion}
	\begin{align}
		\frac{d \ParticlePosition}{dt} & =  
		 \FlowVelocity (\ParticlePosition, t) + \SwimmingVelocity \, \SwimmingDirection, \label{eq:Xdot_motion}\\
		\SwimmingDirection(t) & = \ControlDirection(t) .\label{eq:pdot_motion}
	\end{align}
\end{subequations}

We assume that the plankter senses the local flow velocity gradient $\Gradients$ and the vertical direction $\Direction$. 
It responds to this information by choosing a  direction $\ControlDirection(\Gradients,\Direction)$, without any memory.
The metric used to quantify the performance of the plankters is the effective velocity, $\Performance*$, defined as the long-time average velocity along $\Direction$
\begin{equation}
    \label{eq:performance}
    \Performance* = \lim_{T\to\infty} \frac{\ParticlePosition (T) - \ParticlePosition (0)}{T} \cdot \Direction.
\end{equation}
In the language of control theory (resp. reinforcement learning), $\ControlDirection(\Gradients,\Direction)$ is the control (resp. policy) and $\Performance*$ is the objective function (resp. return).

Using a Taylor expansion of $\FlowVelocity (\vec{x}, t)$ in the neighborhood of the current time $t_0$ and position $\ParticlePosition_0 = \ParticlePosition (t_0)$, the velocity field can be approximated as
\begin{equation}
    \label{eq:linear}
    \FlowVelocity (\vec{x}, \TimeHorizon) \approx \FlowVelocity_0 + (\Gradients)_0 \cdot \left(\vec{x}  - \ParticlePosition_0\right)+ \, \left(\frac{\partial \FlowVelocity}{\partial t} \right)_0 \left( \TimeHorizon - t_0\right),
\end{equation}
where the subscript 0 indicates a variable evaluated at time $t_0$ and location $\ParticlePosition_0$ (e.g., $\FlowVelocity_0 = \FlowVelocity(\ParticlePosition_0, t_0)$).

Inserting approximation \eqref{eq:linear} into Eqs. \eqref{eq:motion} and integrating \footnote{See Supplemental Material for additional details and results\label{fn:supmat}.}, 
one can show that the displacement along $\Direction$ between time $t_0$ and  $t_0+\TimeHorizon$ is maximized when
\begin{equation}
    \label{eq:optimal_swimming_direction_partial}
    \ControlDirection*(t_0+t) = \left[ \exp \left( \left( \TimeHorizon - t \right) \Gradients*_0 \right) \right]^T \cdot \Direction,
\end{equation}
with $0 \leq t \leq \tau$,  $\exp(\cdot)$  the matrix exponential, and $[\cdot]^{T}$ denoting the transpose.
For a plankter continuously sensing the flow, we can set $t=0$ and drop the subscript 0.
After normalization, the preferred direction of the surfing strategy is thus
\begin{equation}
    \label{eq:optimal_swimming_direction}
    \ControlDirectionOpt = \frac{\ControlDirectionOpt*}{\norm{\ControlDirectionOpt*}}, \quad \text{with } \ControlDirectionOpt* = \left[ \exp \left( \TimeHorizon \Gradients \right) \right]^T \cdot \Direction,
\end{equation}
with the time horizon $\TimeHorizon$ a free parameter of this surfing strategy.

The fully-turbulent flow that models the plankter environment is obtained from the Johns Hopkins Turbulence Database \cite{li2008public, perlman2007data}. 
It is a direct numerical simulation of a 3D homogeneous isotropic turbulent flow  with $\mathit{Re}_\lambda = 418$.
The Lagrangian equations of plankter motion, Eqs. \eqref{eq:motion}, are integrated with an in-house open-source code, SHELD0N \footnote{Our in-house code is available at \url{http://www.github.com/C0PEP0D/sheld0n}.}, 
using a 4th-order temporal Runge-Kutta scheme and a 6th-order spatial interpolation scheme.

In a turbulent flow, the smallest flow features are described by the Kolmogorov time $\KolmogorovT$ and Kolmogorov velocity $\KolmogorovU$
\begin{equation}
    \label{eq:kolmogorov}
    \KolmogorovT = (\nu / \epsilon)^{1/2}, \quad \KolmogorovU = (\nu \epsilon)^{1/4},
\end{equation}
with $\nu$ the kinematic viscosity and $\epsilon$ the average dissipation rate \cite{frisch1995turbulence}.
The largest flow features are characterized by the large-eddy turnover time $T_L$ and the root-mean-square velocity $u_\mathrm{rms}$, with $T_L \approx 47 \KolmogorovT$ and $u_\mathrm{rms} \approx 10 \KolmogorovU$ here.
Unless mentioned otherwise, the performance is evaluated after a time $T \approx 5 T_L$, using Eq. \eqref{eq:performance}, and averaged over $N$ plankters with random initial positions.
Averaged quantities are noted $\left\langle \cdot \right\rangle$.
$N$ varies from $10$ for $\SwimmingVelocity = 20 \KolmogorovU$ to $16384$ for $\SwimmingVelocity = \KolmogorovU/2$ to ensure similar uncertainties on performance.
\begin{figure}%[H]
    \centering
    \includegraphics{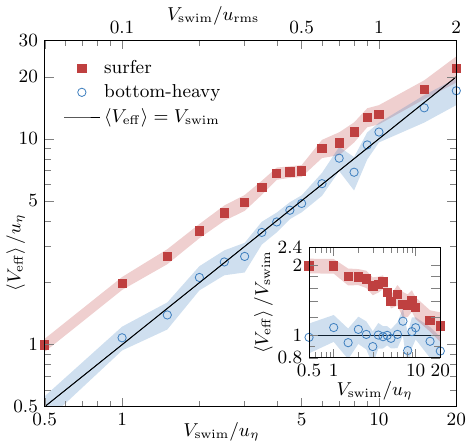}
    \caption{
        Effective upward velocity ($\Performance*$, defined by Eq. \eqref{eq:performance}) as a function of the swimming velocity ($\SwimmingVelocity$) for a surfer ($\ControlDirection=\ControlDirectionOpt$ with optimal time horizon $\TimeHorizon=\TimeHorizonOpt$) and for a bottom-heavy swimmer ($\ControlDirection=\Direction$), here assuming instantaneous reorientation time (Eq. \ref{eq:motion}). 
        Velocities are normalized either by the Kolmogorov velocity ($\KolmogorovU$, Eq. \eqref{eq:kolmogorov}) (bottom x-axis) or by the root-mean-square velocity $u_\mathrm{rms}$ (top x-axis). 
        The inset presents the same data where the effective upward velocity is normalized by the swimming velocity.
        The solid line represents $\Performance* = \SwimmingVelocity$. Shaded areas correspond to 95\% confidence intervals.
    }
    \label{fig:results}
\end{figure}

We now assess the performance of surfers, which actively choose a preferred direction $\ControlDirection=\ControlDirectionOpt$, given by Eq. \eqref{eq:optimal_swimming_direction}.
For that purpose, we compare them to bottom-heavy swimmers, which passively align upwards, that is $\ControlDirection=\Direction$.
In Fig. \ref{fig:results}, we show that surfers  can reach effective speeds, $\Performance*$, as large as twice their swimming speed when $\SwimmingVelocity \lesssim \KolmogorovU$.
They systematically outperform bottom-heavy swimmers, whose performance is $\Performance* = \SwimmingVelocity$ in the limit of instantaneous reorientation (Eq. \ref{eq:pdot_motion}). 
This is because turbulence acts as a random noise of zero mean for bottom-heavy swimmers. In contrast, surfers can exploit the turbulent flow by biasing the sampling of vertical flow velocities \cite{Note1}.
This shows that sensing flow gradients is beneficial for navigation in turbulence  and that surfing allows to exploit this information.

\begin{figure}%[H]
    \centering
    \includegraphics{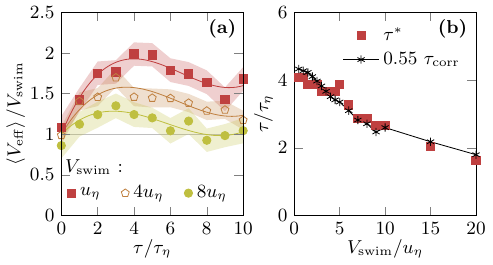}
    \caption{
        Influence of the time horizon on the surfing strategy.
        (a)  Effect of the time horizon ($\TimeHorizon$, Eq. \eqref{eq:optimal_swimming_direction}) on the effective velocity ($\Performance$, Eq. \eqref{eq:performance}), for different swimming velocities $\SwimmingVelocity$. 
        Shaded area represents the 95\% confidence interval. 
        Solid lines represent a fit with Chebyshev polynomials of degree 3.
        (b) Correlation time $\CorrelationTime$, defined in Eq. \eqref{eq:corr}, and optimal time horizon $\TimeHorizonOpt$ as a function of swimming velocity ($\TimeHorizonOpt$ is evaluated using the fitted polynomial).
    }
    \label{fig:more}
\end{figure}

To determine the optimal value of the time horizon $\TimeHorizonOpt$, we look numerically for the best performance when $\TimeHorizon$ varies in the range $[0, 10\KolmogorovT]$  (Fig.~\ref{fig:more}a).
For all swimming velocities $\SwimmingVelocity$, the performance $\Performance$ has a clear velocity-dependent maximum at $\TimeHorizonOpt(\SwimmingVelocity) = O(\KolmogorovT)$.
When $\TimeHorizon \ll \KolmogorovT$, surfers do not use gradient sensing and swim upwards (see Eq. \ref{eq:optimal_swimming_direction}). 
Acting as bottom-heavy swimmers, their performance is $\Performance* = \SwimmingVelocity$.
When $\TimeHorizon \gg \KolmogorovT$, the steady linear approximation of the flow, given in Eq. \eqref{eq:linear}, breaks down and the  planned route becomes irrelevant.
The optimal value $\TimeHorizonOpt$ can thus be interpreted as the time interval over which the steady linear approximation of the flow is reasonable.
For $\SwimmingVelocity = \KolmogorovU$, the optimal time horizon is $\TimeHorizonOpt \approx 4 \KolmogorovT$.
Although our results are based on a single simulation at large Reynolds number, we expect our conclusions to be qualitatively independent of $\mathit{Re}$, because of the universality of turbulence at small scale in the limit of large $\mathit{Re}$ \cite{frisch1995turbulence}.

The relative surfer performance, $\Performance / \SwimmingVelocity$, decreases as the swimming speed increases (Fig.~\ref{fig:results}).
This is because the correlation time of the flow gradients measured by a plankter decreases as $\SwimmingVelocity$ increases. In other words, when plankters swim faster, the surrounding flow changes faster. 
Therefore, $\TimeHorizonOpt$ and $\Performance$  decrease with swimming speed.
Supported by this observation, we hypothesize that the optimal time horizon $\TimeHorizonOpt$ scales as a correlation time $\CorrelationTime$. 
We define $\CorrelationTime$ as the integral of the period, $2\pi/\omega$, weighted by the spectrum of $\mathrm{Tr} ( [ \Gradients ]^2 )^{1/2}$ measured along trajectories of plankters
\begin{equation}
    \label{eq:corr}
    \CorrelationTime (\SwimmingVelocity) = \frac{\int \left\langle I(\omega) \right\rangle \frac{2\pi}{\omega} \, d\omega}{\int \left\langle I(\omega) \right\rangle \, d\omega},
\end{equation}
where $I(\omega)$ is the temporal modulus of the Fourier transform of $\mathrm{Tr} ( [ \Gradients ]^2 )^{1/2}$ and depends on the swimming velocity. 
Figure \ref{fig:more}b shows that, up to a multiplicative constant, $\CorrelationTime$ is a good predictor of the optimal time horizon with $\TimeHorizonOpt \approx 0.55 \CorrelationTime$.
The choice of $I$ in Eq. \ref{eq:corr} is not unique, but other invariants of the velocity gradient yield similar results \cite{Note1}.

\begin{figure}%[H]
    \centering
    \includegraphics{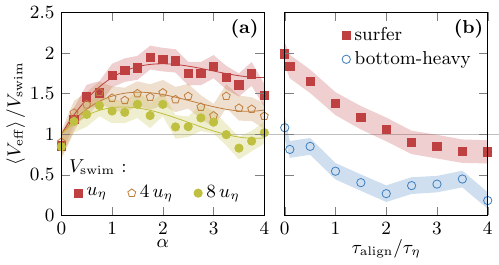}
    \caption{
    (a) Performance ($\Performance*/\SwimmingVelocity$) of surfers using a time dependent $\TimeHorizon$ as a function of the dimensionless constant $\alpha$ (Eq. \eqref{eq:tau_adapt}) for different swimming velocities $\SwimmingVelocity$.
    %Shaded area represents the 95\% confidence interval. 
    Solid lines represent a fit with Chebyshev polynomials of degree 3.
    (b) $\Performance*/\SwimmingVelocity$ of surfers and bottom-heavy swimmers as a function of the reorientation time $\ReorientationTime$. $\SwimmingVelocity = \KolmogorovU$.
    Shaded area represents the 95\% confidence interval. 
    }
    \label{fig:effects}
\end{figure}
We now discuss the applicability of the surfing behavior to more realistic situations relevant to planktonic navigation.
First, the turbulence intensity of plankton environments fluctuates on short time scales \cite{franks2022oceanic}.
This may appear as a problem since surfers need to evaluate the value of $\KolmogorovT$ of their local environment to choose the optimal time horizon $\TimeHorizonOpt$. 
But in practice $\KolmogorovT$ can be estimated from the velocity gradient itself since $\KolmogorovT \sim 1/\abs{\mathrm{sym} \Gradients*}$ where $\abs{.}$ is the Frobenius norm \cite{yu_lagrangian_2010}.
This suggests a refinement of the surfing strategy where $\TimeHorizon$ in Eq. \eqref{eq:optimal_swimming_direction} is replaced by
\begin{equation}
    \label{eq:tau_adapt}
    \TimeHorizon = \frac{\alpha}{\norm{\mathrm{sym} \Gradients*}},
\end{equation}
with $\alpha$ a dimensionless parameter, which can be viewed as a dimensionless time horizon.
In Fig.~\ref{fig:effects}a, we show that surfers using this modified strategy perform as well as surfers with a constant time horizon $\TimeHorizon$. 
For $\SwimmingVelocity = \KolmogorovU$, the optimal value of parameter $\alpha$ is $\alpha^* \approx 2$. 
This value is presumably independent of the turbulence intensity.

Second, sensing and motor control of real organisms may be subject to noise.
We show in Supplemental Material \cite{Note1} that the surfing strategy is robust to noisy measures of $\Direction$ or $\Gradients$ and to noisy control of its preferred direction $\ControlDirection$: for all noise sources, the performance remains practically unchanged for noise up to $25\%$. 

Third, the equations of motion given in Eqs. \eqref{eq:motion} assume an instantaneous reorientation. 
This assumption would require the plankter to exert an infinitely large torque on the fluid. 
For a finite  torque, the equation of orientation, Eq. \eqref{eq:pdot_motion}, should be replaced by \cite{Pedley1992}
\begin{equation}
		\frac{d \SwimmingDirection}{d t}  =  
		\frac{1}{2} \FlowVorticity (\ParticlePosition, t) \times \SwimmingDirection + \frac{1}{2 \ReorientationTime} \left[ \ControlDirection - (\ControlDirection \cdot \SwimmingDirection) \SwimmingDirection \right],\label{eq:pdot_pedley}
\end{equation}
where $\FlowVorticity (\vec{x}, t) = \vec{\nabla} \times \FlowVelocity$ is the flow vorticity, and where $\ReorientationTime$ is a characteristic reorientation time that arises from the balance between the viscous torque and the aligning torque \cite{Note1}.
Figure~\ref{fig:effects}b shows how the performance of surfers ($\ControlDirection=\ControlDirectionOpt$) and bottom-heavy swimmers ($\ControlDirection=\Direction$) decreases as $\ReorientationTime$ increases.
This loss of performance is essentially due to the flow vorticity, which acts as a noise tilting the swimmer away from its preferred direction.
Nevertheless surfers always outperform bottom-heavy swimmers with the same reorientation time. 
Besides, as long as $\ReorientationTime \lesssim 2\KolmogorovT$, the effective speed of surfers $\Performance*$ remains larger that their swimming speed $\SwimmingVelocity$. 

Vertical migration of plankton is essential to ecologically-important activities such as daily migration, dispersal, and larval settlement.
Here, we assess the expected benefit of the surfing strategy over bottom-heaviness for vertical migration in different marine habitats.
To perform this comparison, we use three typical plankters: a copepod, an invertebrate larva, and a dinoflagellate, whose sizes, velocities and reorientation times are given in Tab.~\ref{tab:typical}. 
\begin{table}[t]
    \caption{\label{tab:typical}%
    Characteristics of typical plankters: 
    size $d$ (in mm), 
    swimming velocity $\SwimmingVelocity$ (in mm\,s$^{-1}$), 
    and reorientation time $\ReorientationTime$ (in s).
    The reorientation time depends on the origin of the alignment torque \cite{Note1}. For surfers, this torque is due to active swimming and $\ReorientationTime^{\NameTss*} = d/(3\SwimmingVelocity)$ with $d$ the plankter size. For bottom-heavy swimmers, it is due to gravity and $\ReorientationTime^{\NameStraight*} = 3\nu/(g \delta)$ \cite{Pedley1992} with $g$ the acceleration of gravity and $\delta$ the distance between the center of mass and the geometrical center (we choose $\delta=d/200$, a value typical for zooplankton \cite{jonsson1989vertical, fields1997implications}).
    }
    \begin{ruledtabular}
        \begin{tabular}{ l c c c c }
             & $d$ & $\SwimmingVelocity$ & $\ReorientationTime^{\NameTss*}$ & $\ReorientationTime^{\NameStraight*}$ %\\ 
             %& (mm) & (mm\,s$^{-1}$) & (s$^{-1}$) & (s$^{-1}$) 
             \\[2pt] 
            \hline \\[-5pt]
            copepod & 1 & 3 & 0.1 & 0.008 \\
            invertebrate larva   & 0.2 & 2 & 0.03 & 0.02 \\
            dinoflagellate & 0.03 & 0.3 & 0.03 & 0.2 \\
        \end{tabular}
    \end{ruledtabular}
\end{table}
\begin{figure}%[H]
    \centering
    \includegraphics{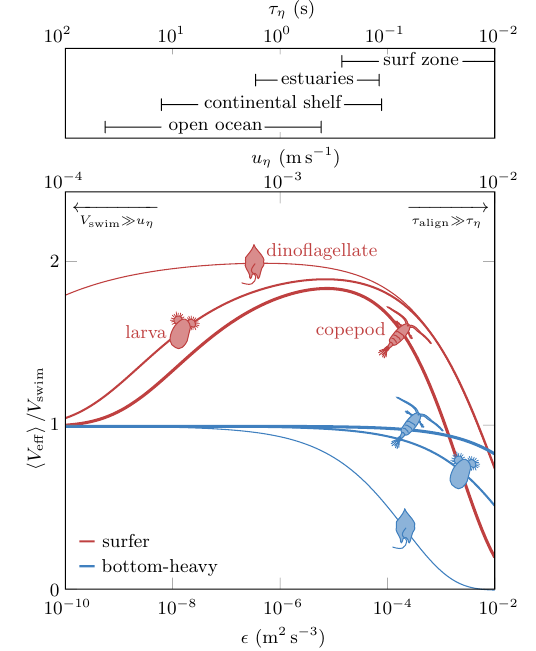}
    \caption{
	Expected vertical migration speed (effective vertical velocity, $\Performance$, Eq. \eqref{eq:performance}, relative to swimming velocity $\SwimmingVelocity$) as a function of the turbulence dissipation rate $\epsilon$ \cite{Note1}.%\footref{fn:supmat}.
       We consider three typical plankters: a copepod, an invertebrate larva, and a dinoflagellate,  whose characteristics are given in Tab. \ref{tab:typical}. 
       Two strategies are compared: the proposed surfing strategy (red) and bottom-heavy swimmers (blue) orienting upwards due to gravity.
       In the upper panel, we indicate the  range of turbulence intensity for different marine habitats (data from \cite{fuchs_seascape-level_2016}) and the corresponding range of  Kolmogorov time $\KolmogorovT$ and Kolmogorov velocity $\KolmogorovU$, Eq. \eqref{eq:kolmogorov}. 
    }
    \label{fig:bio}
\end{figure}
The performance of these typical plankters over a wide range of turbulence conditions \cite{fuchs_seascape-level_2016} is shown in Fig.~\ref{fig:bio}. This figure uses empirical fits deduced from our simulations \cite{Note1} that accounts for the performance drop when $\SwimmingVelocity \gtrsim \KolmogorovU$ or when $\ReorientationTime \gtrsim \KolmogorovT$.
Although it has been suggested that oceanic turbulence might be weaker than initially thought \cite{franks2022oceanic}, this figure shows that typical zooplankton species could benefit from the surfing strategy across a wide range of habitats where vertical migration is crucial, in particular continental shelves, estuaries and open oceans.
% Note however that turbulence intensity decrease drastically with depth in open oceans \cite{franks2022oceanic} ($\epsilon \lesssim 10^{-9}$ deeper than 60m) thus restricting advantageous surfing relatively near the surface. %for typical copepods and invertebrate larvae.

It is interesting to compare the proposed surfing strategy to agents trained by reinforcement learning.
In Ref. \cite{Alageshan2020}, a swimming agent was trained to minimize the time to reach a fixed target using a local measure of vorticity in 3D turbulence.
We evaluated the performance of this trained agent on our task by placing the target infinitely far and using a similar turbulence level ($\mathit{Re}_\lambda = 21$). For the same swimming parameters ($\SwimmingVelocity = 1.5 u_\mathrm{rms}$, $\ReorientationTime = 0.5 \KolmogorovT$), surfers are able to perform 1.5 times better than agents trained by reinforcement learning \cite{Note1}.
Although the comparison remains indicative as reinforcement learning agents were trained with a slightly different objective, it shows that the surfing strategy is performant and should be used as a reference in reinforcement learning problems.

In summary, we have shown that the planktonic navigation problem of going upwards has an approximate analytical solution, which we called surfing. 
The proposed surfing strategy has three important properties:
(1) it is efficient, the effective upward velocity being as large as twice the swimming speed, 
(2) it is adaptive to different turbulent intensities,
(3) and it is robust to finite-time reorientation and  various sources of noise.
We showed that surfing involves a single adjustable parameter, interpreted as a time horizon and related to the correlation time of the flow gradient seen by the swimmer in turbulence.
Finally, we have shown that surfing, which exploits information provided by local velocity gradients, provides a clear benefit over bottom-heaviness for vertical migration of various planktonic species across a wide range of marine habitats.

\begin{acknowledgments}
This project has received funding from the European Research Council (ERC) under the European Union's Horizon 2020 research and innovation program (grant agreement No 834238).
C.E. has been partially funded by a Fulbright Research Scholarship.
M.K. is funded by the National Science Foundation, USA (grant No. IOS-1655318).
Results use data from the Johns Hopkins Turbulence Database.
Centre de Calcul Intensif d’Aix-Marseille is acknowledged for granting access to its high-performance computing resources. 
\end{acknowledgments}

%\bibliographystyle{apsrev4-1}
%\bibliography{bib}% Produces the bibliography via BibTeX.

%

\end{document}

\preprint{APS/123-QED}

\title{Supplemental Material for ``Surfing on turbulence''}

\author{Rémi Monthiller}
\author{Aurore Loisy}
\affiliation{
    Aix Marseille Univ, CNRS, Centrale Marseille, IRPHE, Marseille, France
}
\author{Mimi A. R.  Koehl}
\affiliation{
    Department of Integrative Biology, University of California, Berkeley, CA 94720-3140, USA
}
\author{Benjamin Favier}
\author{Christophe Eloy}
\affiliation{
    Aix Marseille Univ, CNRS, Centrale Marseille, IRPHE, Marseille, France
}

\maketitle

\section{Detailed derivation of the surfing strategy}\label{sec:integration}

To derive the swimming direction chosen with the surfing strategy, we start from the equation of motion of inertialess active particles
\begin{equation}
    \stepcounter{equation}\tag{S{\theequation}}
    \label{eq:s_motion}
    \frac{d \ParticlePosition}{dt} = \FlowVelocity (\ParticlePosition, t) + \SwimmingVelocity \, \ControlDirection \quad \mathrm{and} \quad \SwimmingDirection(t) = \ControlDirection(t) \mathrm{.}
\end{equation}
where $\SwimmingVelocity$ is the swimming velocity, assumed constant, and $\ControlDirection = \SwimmingDirection$ is the time-dependent swimming direction which we want to optimize.
The fluid flow $\FlowVelocity$ at position $\ParticlePosition$ and time $\TimeHorizon$ can be approximated using the following linear approximation
\begin{equation}
    \stepcounter{equation}\tag{S{\theequation}}
    \label{eq:s_linear}
    \FlowVelocity (\vec{x}, \TimeHorizon) \approx \FlowVelocity_0 + (\Gradients)_0 \cdot \left(\vec{x}  - \ParticlePosition_0\right)+ \, \left(\frac{\partial \FlowVelocity}{\partial t} \right)_0 \left( \TimeHorizon - t_0\right).
\end{equation}
Without loss of generality, we can assume $t_0 = 0$ and $\ParticlePosition_0 = \vec{0}$ and drop the $0$ subscript in the following. By substituting this expression in Eq. \eqref{eq:s_motion}, we obtain the linear approximation
\begin{equation}
    \stepcounter{equation}\tag{S{\theequation}}
    \label{eq:s_linear_motion}
     \frac{d \ParticlePosition(\TimeHorizon)}{d \TimeHorizon} \approx \FlowVelocity + \Gradients* \cdot \ParticlePosition(\TimeHorizon)
     + \, \left(\frac{\partial \FlowVelocity}{\partial t} \right) \TimeHorizon + \SwimmingVelocity \,\ControlDirection(\TimeHorizon).
\end{equation}

Integrating this first-order differential equation leads to the following solution for the displacement
\begin{multline}
    \stepcounter{equation}\tag{S{\theequation}}
    \label{eq:s_integration}
    \ParticlePosition(\TimeHorizon) =
    \left[ \exp \left( \TimeHorizon \Gradients* \right) - \matr{Id} \right] \cdot \Gradients*^{-1} \cdot \left[ \FlowVelocity \, + \Gradients*^{-1} \cdot \left(\frac{\partial \FlowVelocity}{\partial t} \right) \, \right] \\
    - \, \TimeHorizon \Gradients*^{-1} \cdot \left(\frac{\partial \FlowVelocity}{\partial t} \right)
    + \, \SwimmingVelocity \int_{0}^{\TimeHorizon} \exp \left( (\TimeHorizon - t) \Gradients \right) \cdot\ControlDirection(t) \, dt ,
\end{multline}
with $\matr{Id}$, the identity matrix.
The problem consists in finding the control $\ControlDirection$ such that the displacement along $\Direction$ is maximum, that is
\begin{equation}
    \stepcounter{equation}\tag{S{\theequation}}
    \text{Find} ~ \ControlDirection ~ \text{such that} ~ \ParticlePosition(\TimeHorizon) \cdot \Direction ~ \text{is maximum}.
\end{equation}
In Eq. \eqref{eq:s_integration}, since $\ControlDirection$ only appears in the last term of the equation, this problem reduces to
\begin{equation}
    \stepcounter{equation}\tag{S{\theequation}}
    \text{Find} ~ \ControlDirection ~ \text{such that} ~ \int_{0}^{\TimeHorizon} \exp \left( (\TimeHorizon - t) \Gradients*_0 \right) \cdot\ControlDirection(t) \cdot \Direction \, dt ~ \text{is maximum}.
\end{equation}
This is done by maximizing the integrand, which can be rewritten in a convenient form
\begin{equation}
    \stepcounter{equation}\tag{S{\theequation}}
    \label{eq:s_inside}
    \text{Find} ~ \ControlDirection ~ \text{such that} ~ \left[ \exp \left( (\TimeHorizon - t) \Gradients*_0 \right) \right]^T \cdot\Direction \cdot \ControlDirection(t) ~ \text{is maximum}.
\end{equation}
The solution to Eq. \eqref{eq:s_inside} is that $\ControlDirection$ must be collinear to $\left[ \exp \left( (\TimeHorizon - t) \Gradients*_0 \right) \right]^T \cdot \Direction$, resulting in the surfing strategy
\begin{equation}
    \stepcounter{equation}\tag{S{\theequation}}
    \label{eq:s_optimal_swimming_direction}
    \ControlDirectionOpt = \frac{\ControlDirectionOpt*}{\norm{\ControlDirectionOpt*}}, \quad \text{with} \quad \ControlDirectionOpt* = \left[ \exp \left( (\TimeHorizon - t) \Gradients* \right) \right]^T \cdot \Direction.
\end{equation}
This expression can be further simplified if we assume a continuous measure of flow velocity gradients, and we can set $t=0$
\begin{equation}
    \stepcounter{equation}\tag{S{\theequation}}
    \label{eq:s_optimal_swimming_direction_final}
    \ControlDirectionOpt = \frac{\ControlDirectionOpt*}{\norm{\ControlDirectionOpt*}}, \quad \text{with} \quad \ControlDirectionOpt* = \left[ \exp \left( \TimeHorizon \Gradients* \right) \right]^T \cdot \Direction.
\end{equation}
The only remaining parameter is the horizon time $\TimeHorizon$, which crucially depends on the temporal correlation of the carrying flow.

This analysis shows that, under our hypotheses, $\Gradients$ is the only information needed to locally optimize the plankter trajectory. While some plankters are able to sense the local flow acceleration $\left({\partial \FlowVelocity}/{\partial t} \right)_0$ \cite{fuchs2015hydrodynamic,fuchs2018waves}, Eq. \eqref{eq:s_optimal_swimming_direction_final} shows that it would not be of any direct use for the problem we consider \footnote{Since acceleration and velocity gradients are correlated in turbulent flows, acceleration may still provide indirect information that could be exploited}.

\section{Flow velocity sampled by surfers}\label{sup:pdf}

\begin{figure}%[H]
    \centering
    \includegraphics{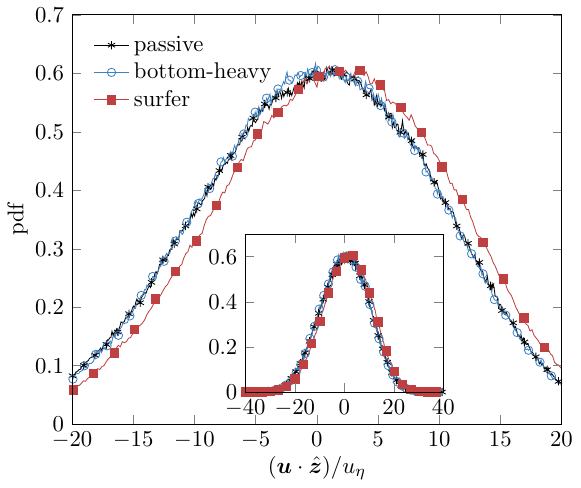}
    \caption{
        Probability density function of the vertical flow velocity sampled along trajectories of passives particles, bottom-heavy swimmers and surfers.
    }
    \label{fig:pdfs}
\end{figure}

In the case of bottom-heavy swimmers, which always swim upwards, turbulence acts as a random noise of zero mean.
In contrast, surfers adapt their response to the local flow gradients, which allow them to preferentially sample upflow velocity regions. 
This bias is illustrated in Fig.~\ref{fig:pdfs}, where we show the distribution of the vertical velocity component of the turbulent flow sampled by surfers, bottom-heavy swimmers, and passive particles.
One can see that the Gaussian distribution is not centered on zero but shifted toward positive values in the case of surfers.

\section{Correlation time of the flow velocity gradients}

The surfing strategy depends on the planning time horizon $\TimeHorizon$.
We expect getting the best performance when this free parameter corresponds to a characteristic correlation time of the flow velocity gradients. 
A shorter value would mean that the information is under-exploited whereas a larger value would mean over-exploiting information of limited accuracy.

\begin{figure}%[H]
    \centering
    \includegraphics{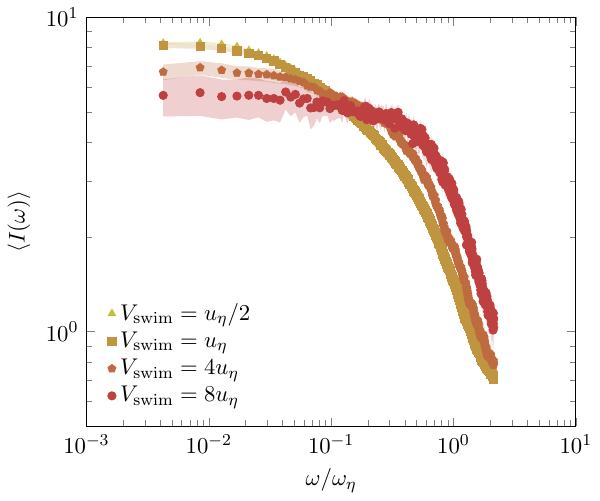}
    \caption{
        Effect of the swimming velocity on the measure of an active particle. $I(\omega)$, module of the temporal Fourier transform of an invariant of the velocity gradients (defined in Eq. \eqref{eq:s_corr}), measured along trajectories of bottom-heavy swimmers ($\ControlDirection = \SwimmingDirection = \Direction$) for various swimming speeds ($\omega_\Kolmogorov = 2 \pi / \KolmogorovT$).
        The shaded area represents the 95\% confidence interval. 
    }
    \label{fig:spect}
\end{figure}

This time does not depend solely on the flow characteristics but also on the plankter velocity. The larger the swimming velocity, the faster it travels through regions of the flow, thus the faster its measure of the flow decorrelates in time.
To highlight this dependence, we plot the module of the temporal Fourier transform of an invariant of the flow velocity gradients
\begin{equation}
    \stepcounter{equation}\tag{S{\theequation}}
    I(\omega) = \abs{ \frac{d}{d\omega} \mathrm{Tr} \left( \left[ \Gradients \right]^2 \right)^{1/2}},
\end{equation}
measured along trajectories of plankters with various swimming velocities (Fig.~\ref{fig:spect}).
As expected, we can see the intensity shifting from low  to high frequencies as the swimming velocity increases.

\begin{figure}%[H]
    \centering
    \includegraphics{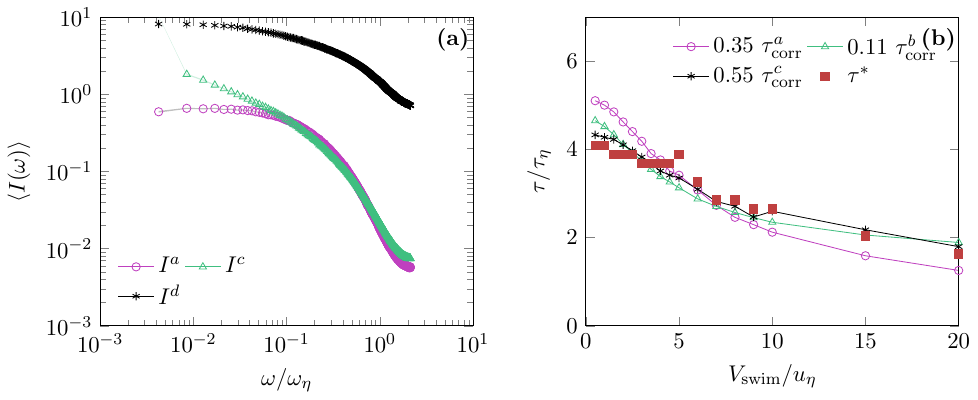}
    \caption{
        Correlation time dependence on the choice of the gradients invariant. (a) $I(\omega)$, module of the temporal Fourier transform of various invariants of the velocity gradients (defined in Tab. \ref{tab:invariants}), measured along trajectories of bottom-heavy swimmers ($\ControlDirection = \SwimmingDirection = \Direction$) for various swimming speeds. 
        $\omega_\Kolmogorov = 2 \pi / \KolmogorovT$.
        The shaded area represents the 95\% confidence interval. 
        (b) Correlation time $\CorrelationTime$ (Eq. \eqref{eq:s_corr}) and optimal time horizon $\TimeHorizonOpt$ as a function of swimming velocity.
    }
    \label{fig:spect_choice}
\end{figure}

From this spectrum, we can introduce a definition of the correlation time of the velocity gradients
\begin{equation}
    \stepcounter{equation}\tag{S{\theequation}}
    \label{eq:s_corr}
    \CorrelationTime (\SwimmingVelocity) = \frac{\int \left\langle I(\omega) \right\rangle \frac{2\pi}{\omega} \, d\omega}{\int \left\langle I(\omega) \right\rangle \, d\omega}
\end{equation}
The choice of the gradient matrix invariant is however not obvious. Any invariant of the gradient matrix could be relevant and further investigation would be needed to find the most relevant invariant to consider.
Figure \ref{fig:spect_choice} shows the differences obtain using the quantities defined in Tab. \ref{tab:invariants}. 
\begin{table}
    \setlength{\tabcolsep}{12pt}
    \begin{tabular}{ ccc  }
        \hline
        $I^a$ & $I^b$ & $I^c$ \\
        \hline
        $\abs{ \frac{d}{d\omega} \Direction \cdot \Gradients \cdot \Direction}$ &
        $\abs{ \frac{d}{d\omega} \norm {\Gradients \cdot \Direction} }$ &
        $\abs{ \frac{d}{d\omega} \mathrm{Tr} \left( \Gradients*^2 \right)^{1/2} }$ \\
        \hline
    \end{tabular}
    \caption{Various possible definitions of $I$ based on different invariants of the flow velocity gradients. $\Direction^{\bot}$ is an arbitrary vector orthogonal to $\Direction$.}
    \label{tab:invariants}
\end{table}
While we observe significant differences between the different invariants, our two most important conclusions are robust: the optimal horizon time $\TimeHorizon$ is always of the order of the Kolmogorov time scale and increasing the swimming velocity tends to reduce this optimal time (due to shorter Lagrangian decorrelation time induced by Doppler-shifting of frequencies).

\section{Estimation of the reorientation time $\ReorientationTime$}\label{sup:time}

Here we derive the reorientation time $\ReorientationTime$ for surfers and bottom-heavy swimmers.
We consider plankters as spheres in a viscous flow. 
The viscous torque exerted on a rotating sphere is \cite{lamb1945hydrodynamics}
\begin{equation}\label{eq.Tmu}
    \stepcounter{equation}\tag{S{\theequation}}
	T_{\mu} = \pi \mu d^3 \omega \mathrm{,}
\end{equation}
with $\mu$ the dynamic viscosity, $d$ the diameter of the plankter and $\omega$ its angular velocity.
The reorientation time can be estimated by balancing this viscous torque with either the active swimming torque for surfers or the passive gravitational torque for bottom-heavy swimmers.

To swim at a constant speed $\SwimmingVelocity$, a microswimmer exerts a force opposite to the Stoke's drag of norm \cite{stokes1851effect}
\begin{equation}
    \stepcounter{equation}\tag{S{\theequation}}
	F_{\mathrm{swim}} = 3 \pi \mu d \SwimmingVelocity \mathrm{,}
\end{equation}
From this force, we can estimate the active torque that such a microswimmer is able to produce by multiplying it by the radius
\begin{equation}
    \stepcounter{equation}\tag{S{\theequation}}
	T_{\mathrm{active}} = \frac{3}{2} \pi \mu d^2 \SwimmingVelocity \mathrm{.}
\end{equation}

Balancing the viscous torque $T_{\mu}$, given in Eq.~\eqref{eq.Tmu}, with the active torque $T_{\mathrm{active}}$ gives a typical angular velocity $\omega_{\mathrm{active}}$. The reorientation time of surfers is then given as
\begin{equation}
    \stepcounter{equation}\tag{S{\theequation}}
	\ReorientationTime^{\NameTss*} = \frac{1}{2\omega_{\mathrm{active}}}=\frac{1}{3} \frac{d}{\SwimmingVelocity}  \mathrm{.}
\end{equation}
We see that this reorientation time only depends on the stride length $\SwimmingVelocity / d$, the number of body lengths traveled per second.

For bottom-heavy swimmers, the gravitational torque is
\begin{equation}
    \stepcounter{equation}\tag{S{\theequation}}
	T_{\NameStraight*} = \frac{1}{6} \pi d^3 \rho g \delta \mathrm{,}
\end{equation}
with $\rho$ the fluid density, $g$ the acceleration of gravity, and $\delta$ the distance between the center of mass and the geometric center.

As for surfers, we can find $\ReorientationTime^{\NameStraight*}$, the reorientation time  of bottom-heavy swimmers, by balancing the gravitational torque with the viscous torque. We find
\begin{equation}
    \stepcounter{equation}\tag{S{\theequation}}
	\ReorientationTime^{\NameStraight*} = 3 \frac{\nu}{g \delta},
\end{equation}
with $\nu$ the kinematic viscosity of the fluid. This expression is identical to the one given by Pedley and Kessler \cite{Pedley1992}.

\section{Robustness to noise}\label{sup:robustness}

Surfers exploit the measure of the upward direction and the velocity gradients to choose a swimming direction. In this section, we will assess how robust is the surfing strategy when there is noise on these 3 quantities: (1) the target direction $\Direction$, (2) the velocity gradients $\Gradients$, and (3) the swimming direction $\SwimmingDirection$.

Robustness to noise is tested by introducing an additive noise with standard deviation $\sigma$ on each of the components of the measure.
For instance, the noisy measure of the target direction is expressed as
\begin{equation}
    \stepcounter{equation}\tag{S{\theequation}}
    \label{eq:noise}
    \Direction_{\textrm{measure}} = \frac{\Direction*_{\textrm{measure}}}{\norm{\Direction*_{\textrm{measure}}}}, \quad \text{with} \quad \Direction*_{\textrm{measure}} = \Direction + \vec{\xi}_{\sigma_{\Direction}} \text{,}
\end{equation}
with $\vec{\xi}_{\sigma}$ a Gaussian white noise so that $\left\langle \vec{\xi}_i(t)\vec{\xi}_j(t') \right\rangle = \sigma^2 \delta_{i,j} \delta(t - t')$.

\begin{figure}%[H]
    \centering
    \includegraphics{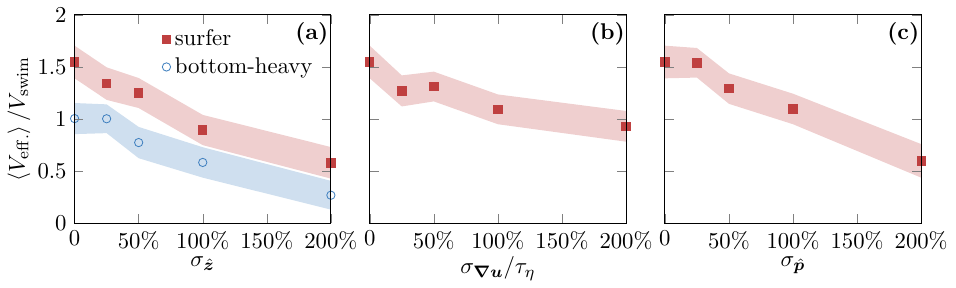}
    \caption{
        Robustness of the surfing strategy. 
        Performance as a function of noise intensity on: 
        (a) the measure of $\Direction$, 
        (b) the flow velocity gradients $\Gradients$ sensing,
        (c) the control of the swimming direction $\SwimmingDirection$.
        The noise is modeled as a Gaussian white noise of standard deviation $\sigma$.
        The shaded area represents the 95\% confidence interval. 
        Parameters: $\SwimmingVelocity = 4 \KolmogorovU$ and $\ReorientationTime = 0$.
    }
    \label{fig:noise}
\end{figure}

Results are summarized in Fig.~\ref{fig:noise}.
We start with the influence of a noisy measure of the target direction (Fig. \ref{fig:noise}a). 
Below $\sigma_{\Direction} = 25\%$, noise has a low impact on performance. 
When noise intensity reaches the magnitude of the signal measured ($\sigma_{\Direction} = 100\%$), performance of surfers and bottom-heavy swimmers decreases significantly.
Real plankters might measure $\Direction$ using either gravity sensing or photoreceptors.
As gravity sensing would be based on a measure of acceleration, we could expect noise to be due to flow acceleration.
Note however that in the ocean, flow acceleration is at it strongest of the order of $0.3$ m.s$^{-2}$ \cite{fuchs_seascape-level_2016}. That corresponds to a noise intensity of $\sigma_{\Direction} = 3\%$ and would not impact performance significantly.

Robustness of the surfing strategy to a noisy measure of $\Gradients$ is illustrated in Fig.~\ref{fig:noise}b.
A noise of intensity lower or equal to $50\%$ leaves the performance essentially unchanged. 
However when the signal is extremely noisy so that the noise magnitude is equal to that of the signal measured ($\sigma_{\Gradients}/\KolmogorovT = 100\%$), performance of surfers decreases significantly.

The effect of a noisy control of the swimming direction $\SwimmingDirection$ is shown in Fig.\ref{fig:noise}c.
Noise intensity has a low impact on performance until it reaches $\sigma_{\SwimmingDirection} = 100\%$ when performance decreases significantly.

Overall, surfing is robust to these various sources of noise: small noise intensities leave the performance essentially unchanged, and the effective speed $\left\langle \Performance \right\rangle$ is greater than the swimming speed $\SwimmingVelocity$ for noise intensities up to $50\%$ of the signal intensity.

\section{Empirical model for typical plankters performance}\label{sup:estimation}

To estimate the performance of typical plankters for a wide range of turbulence conditions (Fig.~5), we fit an empirical model to our numerical data, both for surfers and for bottom-heavy swimmers.

\begin{figure}%[H]
    \centering
    \includegraphics{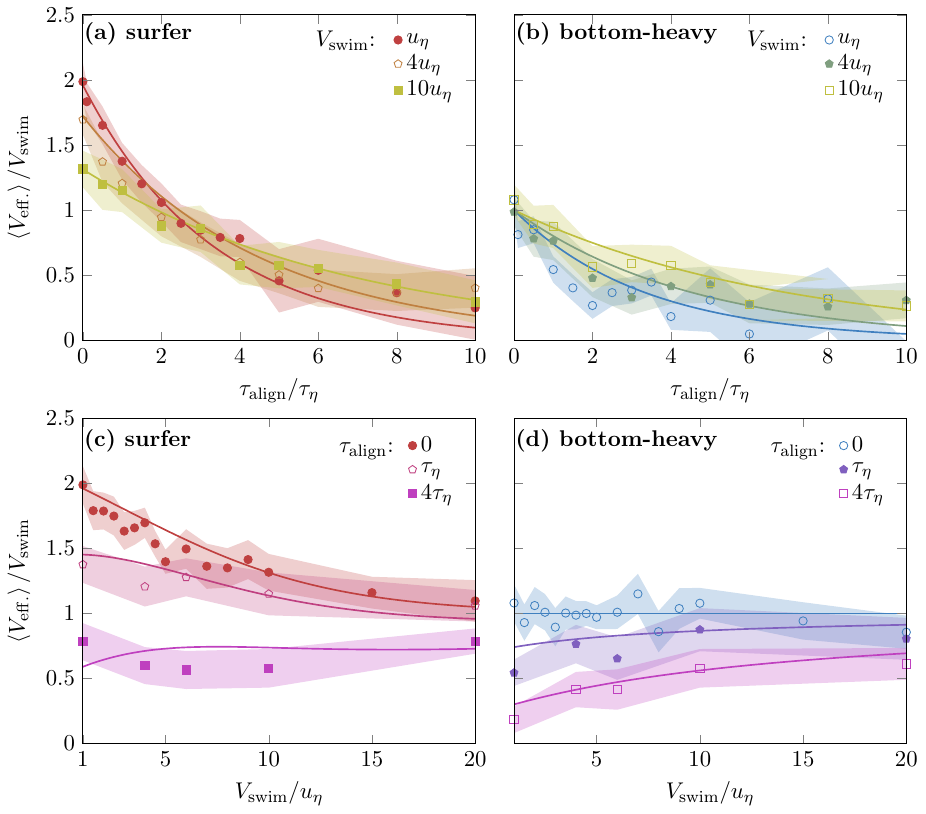}
    \caption{
    	Comparison between the numerical results (symbols) and the empirical fits (solid lines) given in Eqs.~\eqref{eq.fit1} and \eqref{eq.fit2}, for bottom-heavy swimmers and surfers respectively.
	Performance of surfers (a) and bottom-heavy swimmers (b) as a function of $\ReorientationTime$ for different $\SwimmingVelocity$. 
	Performance of surfers (c) and bottom-heavy swimmers (d) as a function of $\SwimmingVelocity$ for different $\ReorientationTime$.
    }
    \label{fig:empirical}
\end{figure}

For bottom-heavy swimmers, we note in Fig.~\ref{fig:empirical}b that $\Performance$ is a function of both $\ReorientationTime$ and $\SwimmingVelocity$. We can interpret these dependencies as follows
Acting as noise, the vorticity tilts the swimming direction of bottom-heavy swimmers away from the vertical. When $\ReorientationTime$ increases, this effect increases too and performance drops.
When $\SwimmingVelocity$ increases, this noise is filtered by swimmers as they sample the flow faster. As a consequence, the effective noise decreases and the performance increases.

We found that the following exponential function could adequately fit our data on the whole range of $\ReorientationTime$ and $\SwimmingVelocity$ studied (Figs.~\ref{fig:empirical}b and \ref{fig:empirical}d)
\begin{equation}\label{eq.fit1}
    \stepcounter{equation}\tag{S{\theequation}}
\frac{\left\langle \Performance \right\rangle}{\SwimmingVelocity} \approx
	 \mathrm{exp} \left( - 0.3 ~ \frac{\ReorientationTime/\KolmogorovT}{0.88 + 0.12 \left( \SwimmingVelocity/\KolmogorovU \right) } \right),\quad
	 \mbox{for bottom-heavy swimmers.}
\end{equation}

For surfers, a similar approach gives the following function (where we assume the same dependence on $\ReorientationTime$ as for bottom-heavy swimmers)
\begin{equation}\label{eq.fit2}
    \stepcounter{equation}\tag{S{\theequation}}
	\frac{\left\langle \Performance \right\rangle}{\SwimmingVelocity} \approx
	 \left[ 1.8 + 0.8 ~ \mathrm{tanh} \left( \frac{3.0 - \left( \SwimmingVelocity/\KolmogorovU \right)}{9.9} \right) \right] ~
	 \mathrm{exp} \left( - 0.3 ~ \frac{\ReorientationTime/\KolmogorovT}{0.88 + 0.12 \left( \SwimmingVelocity/\KolmogorovU \right) } \right)
\end{equation}

The comparison between these empirical fits and the numerical results is shown in Fig. \ref{fig:empirical}. It shows a good agreement that allows to estimate the performance of plankters over a wide range of conditions.

\section{Comparison to reinforcement learning}\label{sup:comparison}

A recent study by Alageshan \textit{et al.} \cite{Alageshan2020} used reinforcement learning to solve a navigation problem similar to ours. In their paper, an agent swims at constant speed in a 3D turbulent flow and seeks to minimize the time to reach a fixed target. The agent uses two observables: the direction towards the target and a local measure of vorticity.
To compare our surfing strategy to the one learned by this agent, we simply replaced their (variable) target direction by our (constant) vertical direction $\Direction$. In other words, we placed the target at infinity.

Their agent was trained in homogeneous isotropic turbulence with $\mathit{Re}_\lambda = 30$. This value is much lower than in the Johns Hopkins Turbulence Database we used ($\mathit{Re}_\lambda = 418$) \cite{li2008public, perlman2007data}. We thus performed a dedicated DNS simulation using the pseudo-spectral, open-source code SNOOPY \cite{lesur2005relevance, lesur2007impact} to generate similar flow conditions.
We solved the Navier-Stokes equations for an in incompressible fluid with kinematic viscosity $\nu = 0.002$ in a tri-periodic domain of size $L = 1$, with a resolution of $128 \times 128 \times 128$.
The flow was made statistically steady thanks to an external forcing delta-correlated in time and localized in spectral space ($3/2 < \norm{\vec{k}} < 5/2$, with $\vec{k}$ the wavevector).
The flow is characterized by a mean dissipation rate $\epsilon = 2 \nu \left\langle \norm{\mathrm{sym} \Gradients}^2 \right\rangle = 0.51$, where $\mathrm{sym} \Gradients$ is the strain rate tensor. The root mean square velocity is $u_{\mathrm{rms}} = 0.42$. The Reynolds number of this flow is comparable to the flow used in Ref. \cite{Alageshan2020}: $\mathit{Re}_\lambda = u_{rms} \lambda/\nu = 21$, where $\lambda = \sqrt{15 \nu u_{rms}^2 / \epsilon} = 0.1$ is the Taylor microscale.

\begin{figure}%[H]
    \centering
    \includegraphics{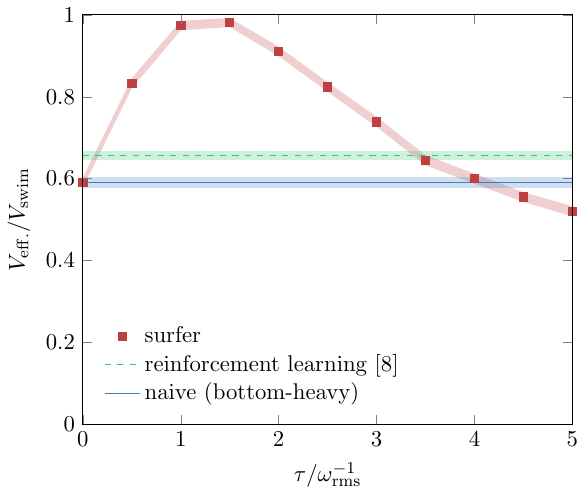}
    \caption{
    	Comparison between the surfing strategy (red symbols), bottom-heavy plankters (blue line) and the ``smart'' strategy obtained by Ref.\cite{Alageshan2020} using reinforcement learning (green dashed line).
   		The performance of surfers is plotted as a function of the time horizon $\TimeHorizon$, which is the only parameter to optimize. The swimmers have the same characteristics: $\SwimmingVelocity = 1.5 \,u_{\mathrm{rms}}$ and $\ReorientationTime = 0.5\, \omega_{\mathrm{rms}}^{-1}$.
    }
    \label{fig:comparison}
\end{figure}

Figure \ref{fig:comparison} shows the performance of surfers as a function of the time horizon $\TimeHorizon$ (the only parameter of our strategy). It is compared to the performance of the strategy obtained by  reinforcement learning \cite{Alageshan2020}.
It shows that surfing  outperforms reinforcement learning: surfers swim 66\% faster than bottom-heavy plankters, while reinforcement learning agents swim 11\% faster than bottom-heavy plankters.
In the original paper \cite{Alageshan2020}, trained agents reach the target 19\% faster than naive agents (which correspond to bottom-heavy plankters). The discrepancy is likely due to a difference in performance metric. 
Surfing not only appears to be a competitive strategy, but it also provides a baseline for future studies using reinforcement learning.

%\bibliography{bib}% Produces the bibliography via BibTeX.

%